\documentclass[pdflatex,sn-aps]{sn-jnl}

\usepackage{graphicx}%
\usepackage{multirow}%
\usepackage{amsmath,amssymb,amsfonts}%
\usepackage{amsthm}%
\usepackage{mathrsfs}%
\usepackage[title]{appendix}%
\usepackage{xcolor}%
\usepackage{textcomp}%
\usepackage{manyfoot}%
\usepackage{booktabs}%
\usepackage{algorithm}%
\usepackage{algorithmicx}%
\usepackage{algpseudocode}%
\usepackage{listings}%
\usepackage{bm}
\usepackage[utf8]{inputenc}

\theoremstyle{thmstyleone}%
\theoremstyle{thmstyletwo}%

\theoremstyle{thmstylethree}%

\graphicspath{
{./image/}
}

\begin{document}

\title[Article Title]{Dimensionality effect on exceptional fermionic superfluidity with spin-dependent asymmetric hopping}


\author*{\fnm{Soma} \sur{Takemori}}\email{takemori.s.041d@m.isct.ac.jp}

\author{\fnm{Kazuki} \sur{Yamamoto},}

\author{\fnm{Akihisa} \sur{Koga}}

\affil{\orgdiv{Department of Physics}, \orgname{Institute of Science Tokyo}, \orgaddress{\street{Meguro}, \city{Tokyo}, \postcode{152-8551}, \country{Japan}}}



\abstract{

  Non-Hermitian (NH) quantum systems host exceptional points (EPs), where eigenstates and eigenvalues coalesce, leading to unconventional many-body phenomena absent in Hermitian systems. While NH fermionic systems with complex interactions exhibit superfluid (SF) breakdown with EPs, spin-dependent asymmetric hopping can stabilize a NH superfluid (NH-SF) that coexists with EPs. In this work, we investigate the quasi-one-dimensional NH attractive Fermi-Hubbard model by using NH BCS theory. We demonstrate that, when the system is regarded as weakly-coupled chains, the exceptional SF phase becomes unstable and metastable (exceptional) SF state appears between the stable SF and normal states. In the one-dimensional limit, the exceptional SF disappear entirely and EPs only appear on the phase boundary between the normal and SF states. These results reveal how dimensional crossover governs the stability of the exceptional SF, providing the insights into the interplay between dimensionality and dissipation, with potential relevance for experimental implications in ultracold atoms.
  }




\maketitle

\section{Introduction}\label{sec_intro}
In recent decades, many-body phenomena in open quantum systems have attracted much attention, and non-Hermitian (NH) systems have been widely explored~\cite{ashida20,fukui98,ashida16,nakagawa18,hamazaki19,yamamoto22,xu20,liu20,yamamoto23sun,yamamoto24}.
One of the important features of NH systems is
the emergence of exceptional points (EPs)
where eigenstates and eigenvalues of the effective Hamiltonian coalesce~\cite{ashida17,lourencco18,zhang20,yang21}.
An NH fermionic system with a complex-valued interaction has been investigated,
and the breakdown of a superfluid (SF) state accompanied by
the emergence of EPs has been clarified~\cite{yamamoto19,kanazawa21,iskin21,he21,takemori24honey,takemori24asym, tajima23topo,li23yang,tajima24,shi24,ji25}.
Remarkably, in Ref.~\cite{takemori25},
a distinct mechanism which stabilizes an NH superfluid (NH-SF) coexisting with EPs
has been proposed by introducing the spin-dependent asymmetric hopping~\cite{uchino12,hayata21sign,yu24,yoshida24} in two- and three-dimensional systems.
Importantly, this phenomenon, referred to as \textit{exceptional SF}, provides
a route to realizing EPs in NH quantum many-body phenomena in ultracold atoms.
Since ultracold gases in optical lattices have been realized and
exhibit characteristic properties in one, two, and three dimensions~\cite{chin06,holten22,sobirey22},
it is important to clarify the role of the dimensionality in stabilizing 
exceptional SF.

To address this issue, we investigate
the quasi-one-dimensional NH attractive Fermi-Hubbard model
with spin-dependent asymmetric hopping.
By employing the NH-BCS theory,
we obtain the superfluid order parameter and associated phase diagram.
In particular, we find that
the exceptional SF becomes unstable
when the system is regarded as the weakly-coupled chains.

The paper is organized as follows. In Sec.~\ref{sec_method}, we introduce the model and obtain the NH gap equation.
We examine the superfluid order parameter and
obtain the phase diagram in Sec.~\ref{sec_result}.
Finally, a conclusion is given in Sec.~\ref{sec_conc}.

\section{Model and method}\label{sec_method}
\begin{figure}[htb]
\centering
\includegraphics[width=0.9\textwidth]{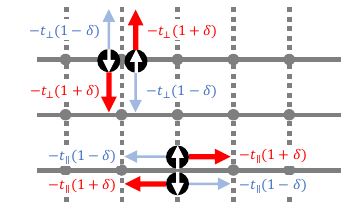}
\caption{
Spin-dependent asymmetric hopping on the quasi-one-dimensional chains.
  }
\label{fig0}
\end{figure}

We consider the quasi-one-dimensional attractive Fermi-Hubbard model with
spin-dependent asymmetric hopping,
which is schematically shown in Fig.~\ref{fig0}.
The Hamiltonian should be given as 
\begin{align}
    H = &-t_{\parallel}\sum_{\langle i,j\rangle_{\parallel},\sigma}[(1+\sigma \gamma)c_{i\sigma}c_{j\sigma} + (1-\sigma \gamma)c_{j\sigma}c_{i\sigma}] \notag\\
    &-t_{\perp} \sum_{\langle i,j\rangle_{\perp},\sigma}[(1+\sigma \gamma)c_{i\sigma}c_{j\sigma}
    + (1-\sigma \gamma)c_{j\sigma}c_{i\sigma}]
    - U\sum_{i}(n_{i\uparrow}-\frac{1}{2})(n_{i\downarrow}-\frac{1}{2}), \label{sdproc_Hami_eq}
\end{align}
where $c_{i\sigma}^{\dagger}(c_{i\sigma})$ is the creation (annihilation) operator of a fermion with spin $\sigma=\uparrow,\downarrow$ at site $i$, and $n_{i\sigma}=c_{i\sigma}^{\dagger}c_{i\sigma}$ is the particle number operator.
$t_{\parallel(\perp)}$ is the hopping amplitude along (between) chains,
$U$ ($>0$) is the interaction strength, and
$\gamma$ $(0\le \gamma\le 1)$ is the rate of asymmetric hopping.
$\langle i,j\rangle_{\parallel(\perp)}$ represents the summation over nearest-neighbor pairs along (between) chains.
The kinetic term of the NH Hamiltonian~\eqref{sdproc_Hami_eq} is rewritten as
\begin{align}
    &H^{\text{kin}} = \sum_{\bm{k}}(\epsilon_{\bm{k}}c_{\bm{k}\uparrow}^{\dagger}c_{\bm{k}\uparrow} + \epsilon_{\bm{k}}^{\ast}c_{\bm{k}\downarrow}^{\dagger}c_{\bm{k}\downarrow}), \\
    &\epsilon_{\bm{k}} = -2t_{\parallel}(\cos k_{\parallel} + i\gamma \sin k_{\parallel}) -2t_{\perp}(\cos k_{\perp} + i\gamma \sin k_{\perp}),
\end{align}
where ${\bm k}=(k_{\parallel}, k_\perp)$ and
$\epsilon_{\bm k}$ is the noninteracting dispersion relation.
In the following, we set $t_{\parallel}=1$ as the unit of energy. 

To analyze the effective ground state of Eq.~\eqref{sdproc_Hami_eq}, we employ the NH-BCS theory~\cite{yamamoto19,takemori25}. We begin with a path-integral representation of the partition function as 
\begin{align}
    &Z = \int \mathcal{D}\bar{c}\mathcal{D}c \exp(-S),\\
    &S = \int_{0}^{\beta'} \sum_{i,\sigma}\bar{c}_{i\sigma}\partial_{\tau}c_{i\sigma} + H(\bar{c}_{i\sigma},c_{i\sigma}),
\end{align}
where $\beta'$ is a statistical weight parameter for the partition function, $\tau$ is the imaginary time, and $H(\bar{c}_{i\sigma}, c_{i\sigma})$ is obtained by replacing the fermion operators with Grassman variables in Eq.~\eqref{sdproc_Hami_eq}. 
Then, by employing the Hubbard-Stratonovich transformation and integrating out the fermionic degrees of freedom, we obtain the NH gap equation from the saddle point of the action as 
\begin{align}
    &\frac{1}{U} = \sum_{\bm{k}} \frac{1}{2E_{\bm{k}}} = \iint d\text{Re}\epsilon\:d\text{Im}\epsilon \:\frac{D_{\gamma}(\epsilon)}{2\sqrt{\epsilon^{2}+\Delta^{2}}}, \label{sdproc_gapeq} \\
    &D_{\gamma}(\epsilon)= \frac{1}{N}\sum_{\bm{k}}\delta(\text{Re}\epsilon - \text{Re}\epsilon_{\bm{k}})\delta(\text{Im}\epsilon - \text{Im}\epsilon_{\bm{k}}), \label{sdproc_effDOSeq}
\end{align}
where $\Delta$ is the order parameter, $E_{\bm{k}}=\sqrt{\epsilon_{\bm{k}}^{2}+\Delta^{2}}$, $\delta(x)$ is the delta function, and $D_{\gamma}(\epsilon)$ is the effective density of states (DOS) on a complex energy plane~\cite{takemori25}.
Note that $D_{\gamma}(\epsilon)=D_{\gamma=1}(\text{Re}\epsilon+i\text{Im}\epsilon/\gamma)$.
\begin{figure}[htb]
\centering
\includegraphics[width=0.9\textwidth]{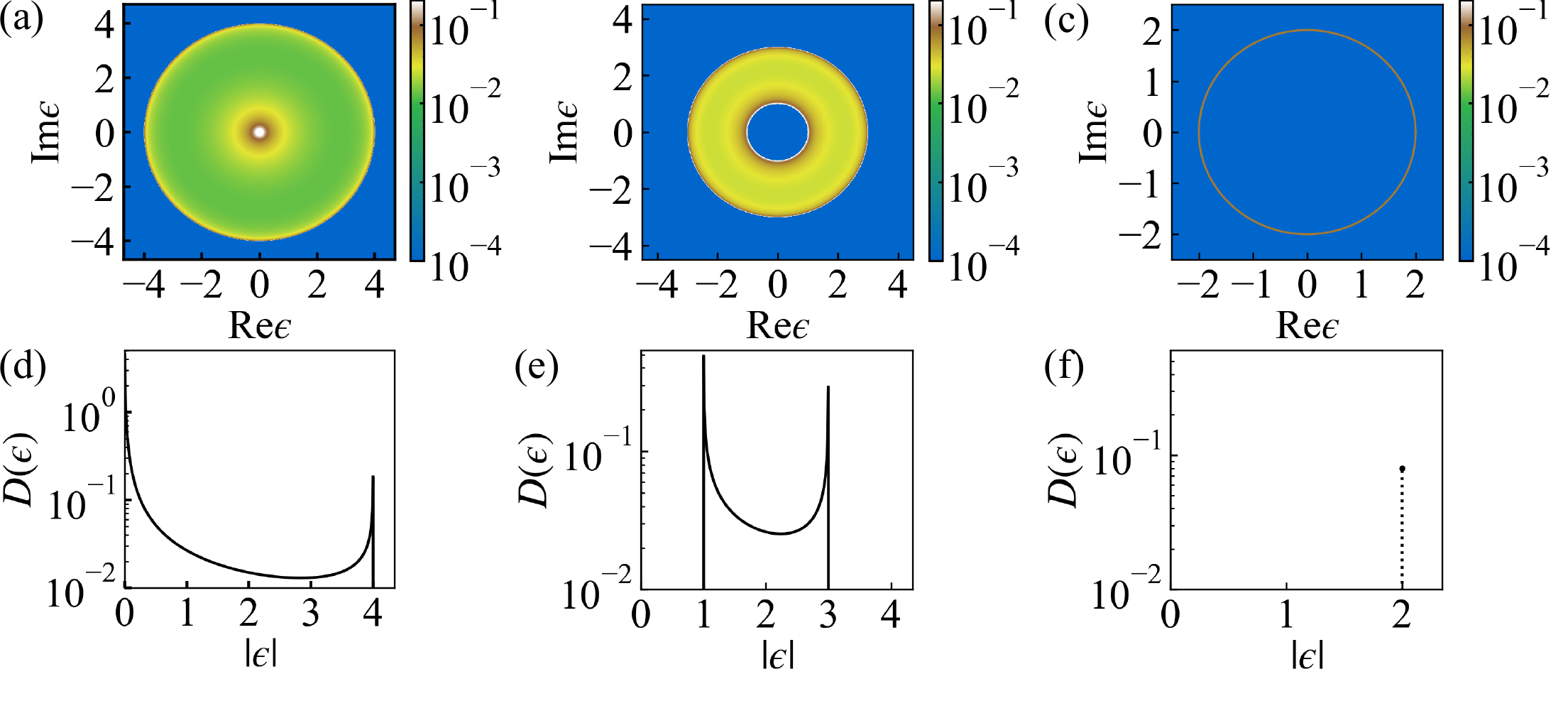}
\caption{Contour plots of the effective DOS for $\gamma=1$: (a) $t_{\perp}=1$, (b) $t_{\perp}=0.5$, and (c) $t_{\perp}=0$. The corresponding cross sections are shown in (d)-(f), respectively.
}
\label{sdproc_DOS_image}
\end{figure}

Figure~\ref{sdproc_DOS_image} shows the effective DOS with $\gamma=1$ for $t_{\perp}=0$, $0.5$, and $1$.
When the two-dimensional system is considered ($t_{\perp}=1.0$),
the effective DOS is finite in the circular region of the complex-$\epsilon$ plane
and diverges at $|\epsilon|=0$, and $4$,
as shown in Figs.~\ref{sdproc_DOS_image}(a) and (d).
For the weakly-coupled chains with $t_{\perp}<t_{\parallel}$, 
the effective DOS appears in a ring-shaped region
in a complex-$\epsilon$ plane and
diverges at $|\epsilon|=2(1\pm t_{\perp})$,
as shown in Figs.~\ref{sdproc_DOS_image}(b) and \ref{sdproc_DOS_image}(e).
As $t_\perp$ decreases, the width of the ring becomes narrower.
In the one-dimensional case with $t_{\perp}=0$,
the effective DOS appears only at $|\epsilon|=2$ as 
\begin{equation}
    D_{\gamma=1}(\epsilon) = \frac{\delta(|\epsilon|-2)}{4\pi}. \label{sdproc_DOS_one_dim_eq}
\end{equation}
An important point is that there exists the region with $D_\gamma(\epsilon)=0$
around $\epsilon=0$ when $t_\perp\neq t_\parallel$.

To discuss the stability of the superfluid state,
we examine the condensation energy, which is given as
\begin{equation}
    E_{\text{cond}} = \frac{\Delta^{2}}{U} -\iint d\text{Re}\epsilon \:d\text{Im}\epsilon \:D_{\gamma}(\epsilon) \left[\sqrt{\epsilon^{2}+\Delta^{2}} - \sqrt{\epsilon^{2}} \right]. \label{sdproc_Ec_eq}
\end{equation}
Note that this condensation energy in our system is always real
although the energy in the NH system takes, in general, a complex value.
When $E_{\text{cond}}<0$, the stable superfluid state is realized,
while for $E_{\text{cond}}>0$, the superfluid state is metastable.
In the following section, we discuss how the SF state is realized
in the quasi-one-dimensional chains.

\section{Superfluid order parameter and phase diagram}\label{sec_result}

\begin{figure}[htb]
\centering
\includegraphics[width=0.9\textwidth]{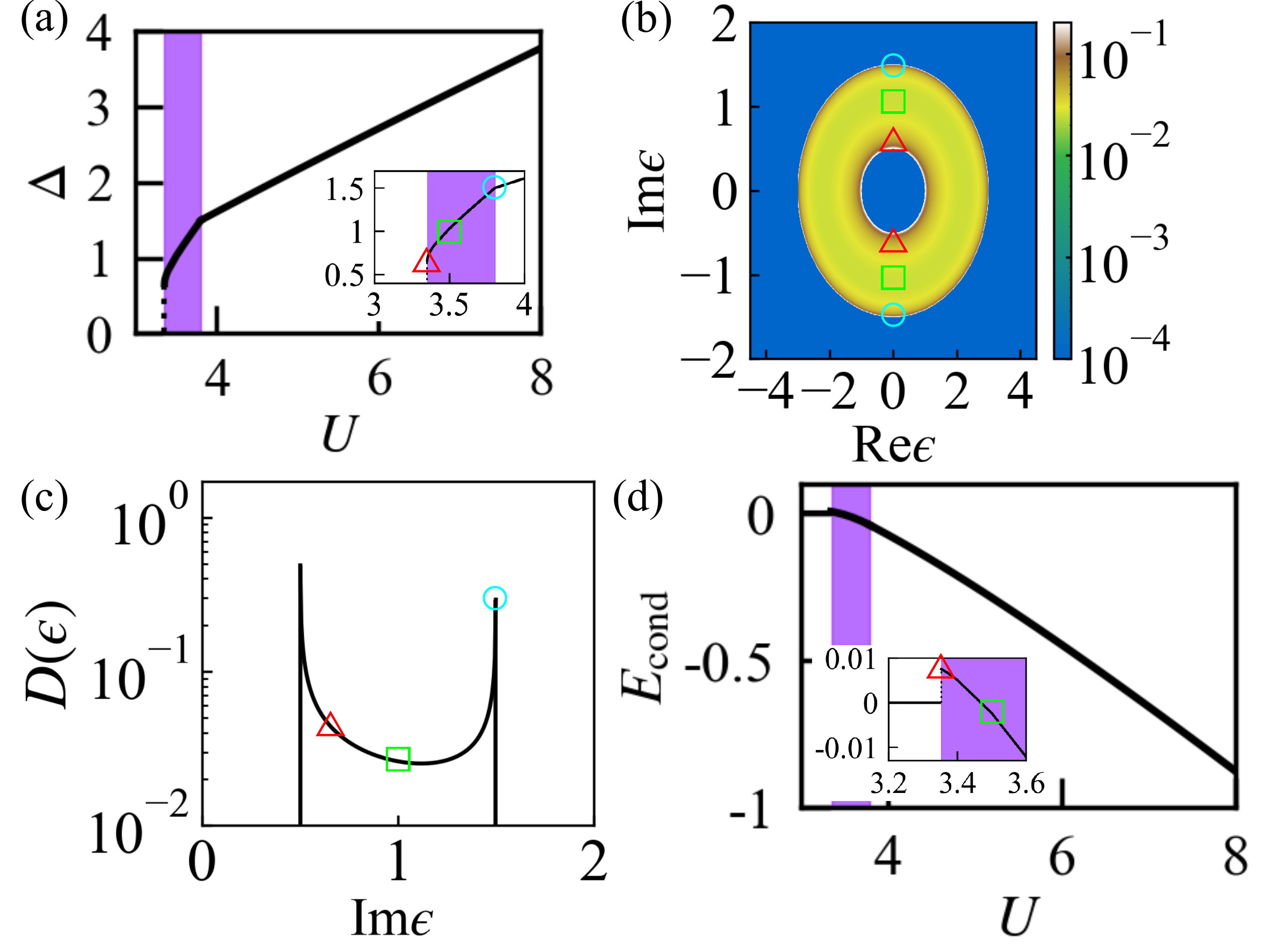}
\caption{(a) Numerical solution of the superfluid order parameter $\Delta$ obtained from the NH gap equation \eqref{sdproc_gapeq}. The exceptional SF emerges in the purple-shaded area. Inset shows the enlarged view near the transition point. (b) Contour plot of the effective DOS~\eqref{sdproc_effDOSeq} and (c) its cross section for $\text{Re}\epsilon=0$. In (b) and (c), the red, green, and blue marks represent EPs given by $\text{Im}\epsilon=\pm\Delta$, where the value of the order parameter is given in (a). (d) Numerical solution of the condensation energy $E_{\text{cond}}$~\eqref{sdproc_Ec_eq}. Inset shows the entangled view near the transition point. The parameters are set to $t_{\perp}=0.5$ and $\gamma=0.5$.}
\label{sdproc_q2d_image}
\end{figure}
In this section, we examine the SF order parameter systematically
to discuss how the interchain hopping affects the stability of the SF state.
We first present the numerical results for $t_{\perp}=0.5$ and $\gamma=0.5$
by solving the NH gap equation~\eqref{sdproc_gapeq}
together with evaluating the condensation energy, 
as shown in Fig~\ref{sdproc_q2d_image}.
For large values of $U$,
the order parameter is finite and $E_{cond}$ is negative.
In this case, $E_{\bm{k}}$ is always finite (not shown) and thereby
the stable NH-SF state is realized without EPs.
As $U$ decreases, the order parameter $\Delta$ monotonically decreases.
When $U\le 3.80$, $\Delta$ is still finite and
EPs emerge at $\epsilon=\pm i\Delta$
in the complex-$\epsilon$ plane.
This implies that the exceptional SF is realized
even in the quasi-one-dimensional system.
EPs appears in the case $3.35\le U\le 3.80$ [see green square in Fig.~\ref{sdproc_q2d_image}],
which means that the exceptional SF emerges in a finite range of attraction $U$.
Further decreasing the attraction,
$\Delta$ suddenly vanishes at $U=3.35$ and 
the normal state is realized for $U<3.35$.
This result is distinct from that in the square lattice case $(t_{\perp}=t_{\parallel})$~\cite{takemori25},
where the SF state is always stable and normal state is never realized.

We note that, at the lower transition point $U=3.35$,
$\Delta=\text{Im}\epsilon$ does not correspond to the lower edge of the effective DOS,
as shown in Fig.~\ref{sdproc_q2d_image}(c).
This contrasts with our previous results~\cite{takemori25}.
The reason is that the physical superfluid gap cannot be obtained
for $\mathrm{Im}\epsilon<0.66$ due to the lack of the effective DOS
for small $|\epsilon|$,
as shown in Figs.~\ref{sdproc_q2d_image}(b) and \ref{sdproc_q2d_image}(c).
Also, the phase transition in quasi-one dimensions is reminiscent of the exceptional SF in three dimensions where the superfluid-normal phase transition occurs~\cite{takemori25}, but the nature of the transitions is different. This is seen in the stability of the superfluid state by numerically evaluating the condensation energy, as shown in Fig.~\ref{sdproc_q2d_image}(d).
When $3.35<U<3.47$,
the condensation energy becomes positive,
indicating that the exceptional SF becomes metastable at finite $U$ in quasi-one dimensional system
-- a feature absent in three dimensions.

\begin{figure}[tb]
\centering
\includegraphics[width=0.9\textwidth]{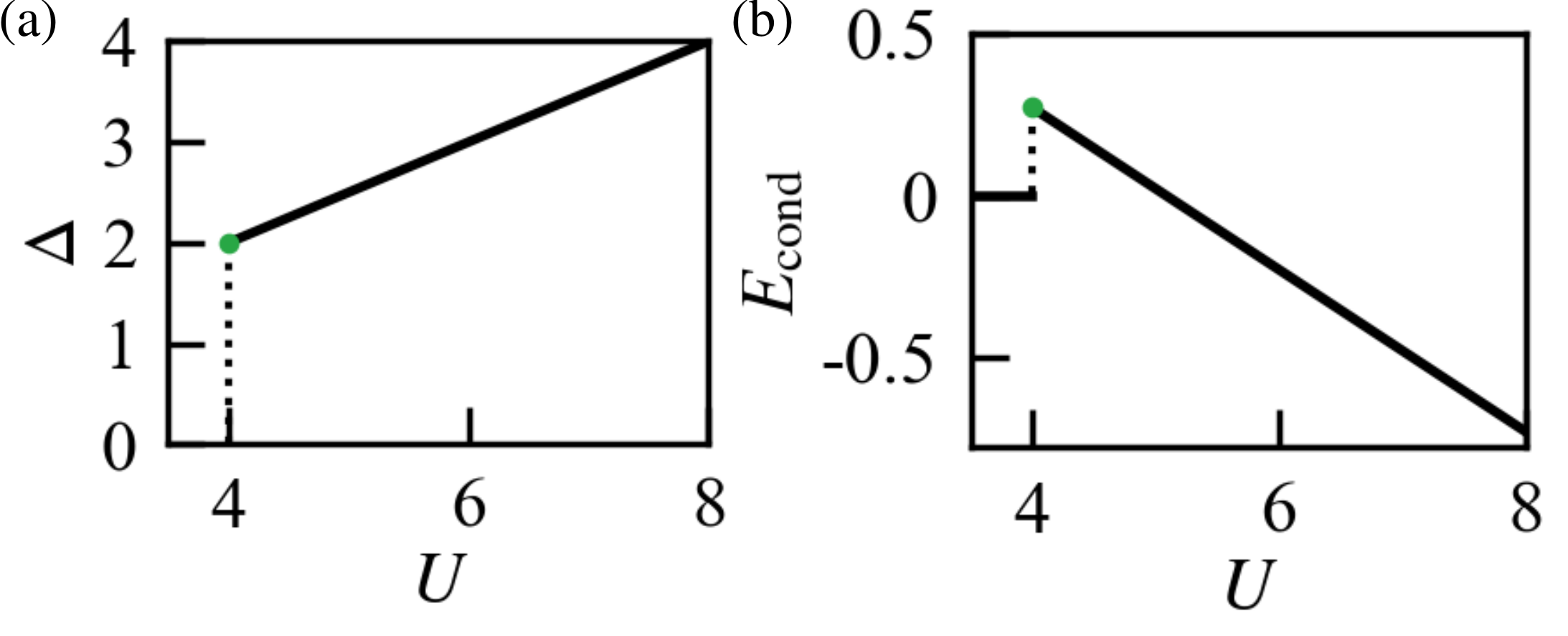}
\caption{(a) Order parameter and (b) condensation energy for $t_{\perp}=0$ and $\gamma=1$. EPs only appear at $U=4$ indicated by a green dot.}
\label{sdproc_1dg1_image}
\end{figure}

Before obtaining the phase diagram, we give 
the analytical solution of the gap equation and
the condensation energy in the limit with $\gamma=1$ and $t_{\perp}=0$.
Using Eq.~\eqref{sdproc_DOS_one_dim_eq} and
integrating the right-hand side of the NH gap equation~\eqref{sdproc_gapeq},
we obtain the self-consistent equation as
\begin{equation}
    \frac{1}{U} = \begin{cases}
        \frac{1}{2\Delta}, & \;\; (\Delta\ge 2), \\
        \frac{1}{2\Delta} -\frac{1}{\pi \Delta}\tan^{-1}(\frac{\sqrt{4-\Delta^{2}}}{\Delta}), & \;\; (\Delta<2).
    \end{cases}
\end{equation}
The condensation energy is then given as
\begin{equation}
    E_{\text{cond}} = \begin{cases}
        -\frac{\Delta}{2} + \frac{4}{\pi}, & \;\; (\Delta\ge 2), \\
        -\frac{\Delta}{2} + \frac{4}{\pi} - \frac{1}{\pi}(2\sqrt{4-\Delta^{2}} -\Delta\tan^{-1}(\frac{\sqrt{4-\Delta^{2}}}{\Delta})), & \;\; (\Delta< 2).
    \end{cases}
\end{equation}
We note that the solution with $\Delta<2$ is unphysical
since it is not adiabatically connected to the solution
in the Hermitian limit $\gamma\to 0$.
In Fig.~\ref{sdproc_1dg1_image}, we show the order parameter $\Delta$ and the condensation energy $E_{\text{cond}}$ as a function of the attraction $U$. When $\Delta=2$, EPs appear with $E_{k}=0$ (see green circle in Fig.~\ref{sdproc_1dg1_image}) only at the phase boundary,
where the order parameter suddenly vanishes.
This is because the effective DOS is finite on the circle with $|\epsilon|=2$.
Moreover, we find that, in Fig.~\ref{sdproc_1dg1_image}, as decreasing the attraction, the condensation energy becomes positive for $U<16/\pi (\sim 5.09)$, indicating that the metastable superfluid state is realized when $4.00<U<5.09$.



\begin{figure}[b]
\centering
\includegraphics[width=0.9\textwidth]{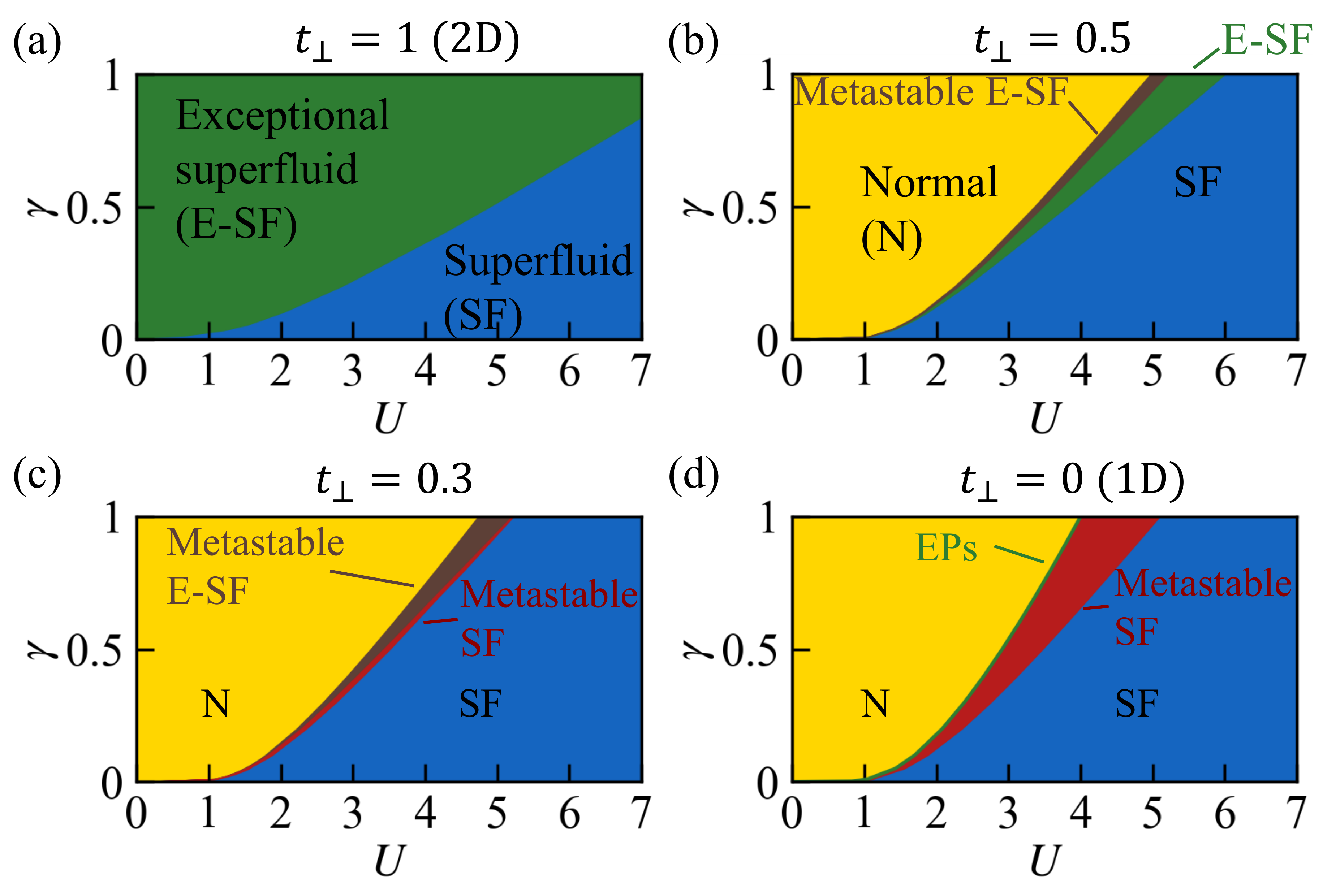}
\caption{Phase diagram of the NH attractive Fermi-Hubbard model with spin-dependent asymmetric hopping in (a) two dimensions with $t_{\perp}=1$, (b), (c) quasi-one dimensions with $t_{\perp}=0.5,0.3$, and (d) one dimensions with $t_{\perp}=0$. 
In (d), EPs only appear on the phase boundary between the normal and the metastable superfluid states.}
\label{sdproc_PD_image}
\end{figure}

Finally, we solve the NH gap equation, evaluate the condensation energy,
and obtain the phase diagrams for $t_\perp=0, 0.3, 0.5$, and $1$,
as shown in Fig.~\ref{sdproc_PD_image}.
It is known that, in the two-dimensional limit with $t_{\perp}=1$,
the superfluid state appears for arbitrary $U$, and
the introduction of asymmetric hopping $\gamma$ induce
the exceptional SF~\cite{takemori25}.
Decreasing $t_{\perp}$, the exceptional SF becomes unstable,
and, instead, the metastable exceptional SF and the normal state
emerge for small $U$.
As $t_{\perp}$ further decreases,
the exceptional SF shrinks, and metastable SF without EPs appears.
In the one-dimensional limit, the (metastable) exceptional SF state never exist
in the phase diagram, as shown in Fig.~\ref{sdproc_PD_image}(d).
We find the metastable SF without EPs between the SF and normal states 
and EPs only appear on the phase boundary.
This phase diagram is reminiscent of the NH-SF with a complex-valued interaction
in three-dimensional optical lattices~\cite{yamamoto19, takemori24asym}.
Thus, the phase diagrams highlight how reducing dimensionality affects the stability of the exceptional SF.

\section{Conclusion}\label{sec_conc}
We have studied the NH attractive Hubbard model with spin-dependent asymmetric hopping on the quasi-one-dimensional systems. We have found that, when the interchain hopping amplitude decreases, the exceptional SF becomes unstable and metastable (exceptional) superfluid state emerges between the stable superfluid and the normal states. Moreover, we have elucidated that the exceptional SF does not appear in the one-dimensional limit. 
As the dissipative fermionic superfluidity in the Lindblad master equation has been proposed recently \cite{yamamoto21}, it is an interesting problem to study how the liouvillian EPs affect the dynamics of collective excitations of fermionic superfluids.

\backmatter

\section*{Declarations}

\begin{itemize}
\item \textbf{Funding:}
 This work was supported by Grant-in-Aid for Scientific Research from JSPS,
KAKENHI Grants No. JP25K17327 (K.Y.), JP22K03525, JP25H01521, and JP25H01398 (A.K.). 
S.T. was supported by the Sasakawa Scientific Research Grant from the Japan Science Society and JST SPRING, Japan Grant Number JPMJSP2180. This work was partly funded by Hirose Foundation, the Precise Measurement Technology Promotion Foundation, and the Fujikura Foundation.
\item \textbf{Conflict of interest/Competing interests:}
 The authors declare no conflict of interest and competing interests.
\item \textbf{Data availability:}
 Data sets generated during the current study are available from the corresponding author upon reasonable request.
\item \textbf{Code availability:}
 The codes used for numerical calculations are available from the corresponding author upon reasonable request.
\item \textbf{Author contribution:}
 All authors contributed to the manuscript writing, the discussion, and the interpretation of the analytical and numerical results. S.T. carried out all the analytical and numerical calculations, and drafted the manuscript. K.Y. and A.K. reviewed the analytical and numerical calculations, suggested improvements, and reorganized the manuscript.
\end{itemize}

\end{document}